\documentclass[traditabstract]{aa}

\usepackage{graphicx}
\usepackage{txfonts}
\usepackage{float}

\usepackage{natbib}
\bibpunct{(}{)}{;}{a}{}{,}

\usepackage{hyperref}
\hypersetup{colorlinks=true,
            citecolor=blue,
            linkcolor=blue}

\usepackage{color}

\def \be {\begin{equation}}
\def \ee {\end{equation}}
\def \omg {\Omega}
\def \vpi {\omega}

\newcommand{\tpr}[1]{\textcolor{red}{#1}}

\def\bfx#1{#1}

\def\figpath{}
\def\bibpath{}
\def \llabel#1{\label{#1}}

\begin{document}


\title{Why do warm Neptunes present nonzero eccentricity?}
\titlerunning{Why do warm Neptunes present nonzero eccentricity?}

\author{
A. C. M. Correia\inst{1,2}
\and
V. Bourrier\inst{3}
\and
J.-B. Delisle\inst{3,2}
}

 
\institute{
CFisUC, Department of Physics, University of Coimbra, 3004-516 Coimbra, Portugal
\and 
ASD, IMCCE, Observatoire de Paris, PSL Universit\'e, 77 Av. Denfert-Rochereau, 75014 Paris, France
\and
Observatoire de l'Universit\'e de Gen\`eve, 51 chemin des Maillettes, 1290 Sauverny, Switzerland
}

\date{Received ; accepted To be inserted later}

\abstract{
Most Neptune-mass planets in close-in orbits (orbital periods less than a few days) present nonzero eccentricity, typically around 0.15. 
This is somehow unexpected, as these planets undergo strong tidal dissipation that should circularize their orbits in a time-scale shorter than the age of the system. 
In this paper we discuss some mechanisms that can oppose to bodily tides, namely, thermal atmospheric tides, evaporation of the atmosphere, and excitation from a distant companion. 
In the first two cases, the eccentricity can increase consistently, while in the last one, the eccentricity can only be excited for a limited amount of time (that may nevertheless exceed the age of the system). 
We show the limitations of these different mechanisms and how some of them could, depending on specific properties of the observed planetary systems, account for their presently observed eccentricities.
}


   \maketitle
%

\section{Introduction}
\llabel{intro}

Planets with orbital periods $P_{\rm orb} \lesssim 10$~day undergo strong tidal interactions with the parent star \citep[e.g.,][]{MacDonald_1964}.
Bodily (or gravitational) tides are raised on the planet by the star because of the gravitational gradient across the planet. 
Since planets are not perfectly rigid, there is a distortion that gives rise to a tidal bulge. 
The dissipation of the mechanical energy inside the planet introduces a delay, and, hence, a phase shift between the initial perturbation and the maximal tidal deformation.
As a consequence, the star exerts a torque on the tidal bulge which modifies the spin and the orbit of the planet.

The ultimate stage for tidal evolution is the synchronization of the rotation and orbital periods, alignment of the planet spin axis with the normal to the orbit (zero planet obliquity), and the circularization of the orbit \citep[e.g.,][]{Hut_1980, Adams_Bloch_2015}.
Although the spin of close-in planets quickly evolves into an equilibrium configuration, the orbital evolution is much slower \citep[e.g.,][]{Correia_Laskar_2010B}.
The circularization characteristic timescale is given by expression (\ref{180501a}) using a constant-$Q$ tidal model
\be
\tau_{\rm circ} = \frac{P_{\rm orb}}{21 \pi} \left( \frac{m}{M} \right) \left( \frac{a}{R} \right)^5 \frac{Q}{k_2} 
\ , \llabel{180702b}
\ee
where $P_{\rm orb}$ is the orbital period, $m$ is the mass of the planet, $M$ is the mass of the star, $a$ is the semi-major axis of the orbit, $R$ is the average radius of the planet, and $k_2$ is the second Love number for potential.
$Q$ is the tidal quality-factor, which measures the ratio of energy dissipated during one period of tidal stress over the peak energy stored in the system during the same period.

\begin{figure}
\centering
    \includegraphics[width=\columnwidth]{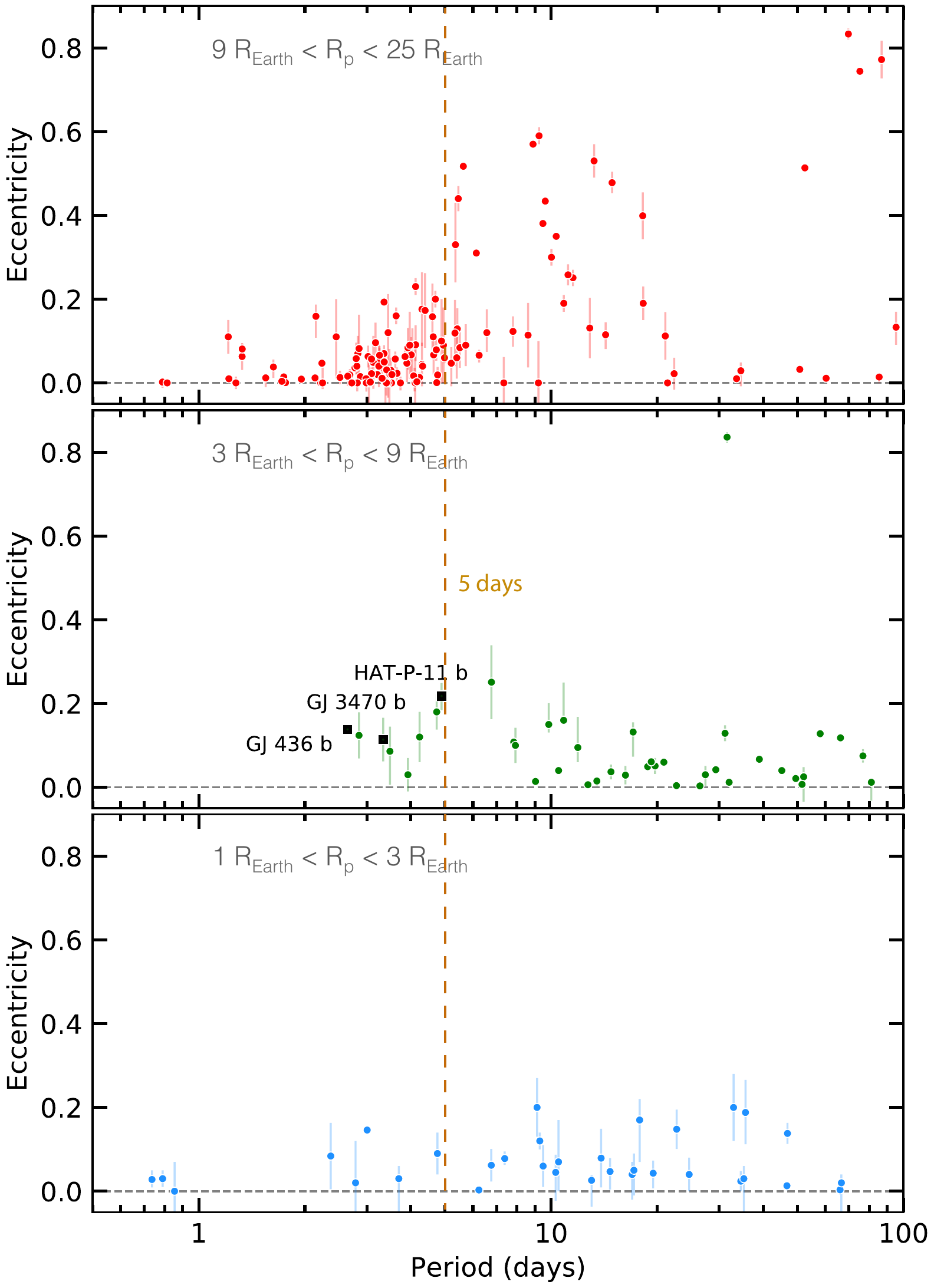}
 \caption{Distribution of eccentricities as a function of orbital period. Only eccentricities measured with uncertainties smaller than 0.1 are shown. Planets are arbitrarily separated in three categories by radius: Jupiter-size (red, top panel), Neptune-size (green, middle panel), and Earth-size (blue, bottom panel). In the middle panel, black squares highlight three iconic planets representative of the warm Neptunes population.} 
\label{HNFig}
\end{figure}

For Uranus and Neptune, $Q/k_2 \sim 10^5$ \citep{Tittemore_Wisdom_1990, Banfield_Murray_1992}.
Assuming that Neptune-mass planets have similar rheologies, we obtain (Eq.\,(\ref{180702b}))
\be
\tau_{\rm circ} \mathrm{[Gyr]} \sim \frac{1}{200} \left( \frac{M}{M_\odot} \right)^{2/3} \left( \frac{P_{\rm orb}}{\mathrm{day}} \right)^{13/3}
\ . \llabel{180702c}
\ee
According to this expression, all Neptune-mass planets with $P_{\rm orb} < 5$~day should circularize in less than 5~Gyr. 
However, observations of Neptune-mass planets such as GJ\,436\,b put this scenario into question. 
In a short-period orbit with $P_{\rm orb} = 2.644$~day \citep{Lanotte_etal_2014} around a cool star with $M = 0.445 \, M_\odot$ \citep{Mann_etal_2015}, GJ\,436\,b has a circularization timescale of $\tau_{\rm circ} \sim 0.2$~Gyr (Eq.\,(\ref{180702c})). 
This is much shorter than the stellar age of about 4 to 8~Gyr \citep{Bourrier_etal_2018b}, yet the planetary orbit has an eccentricity $e = 0.162 \pm 0.004$ \citep{Lanotte_etal_2014}, which is clearly different from zero. 
Several similar cases have been reported in the literature, such as HD\,125612\,c \citep{LoCurto_etal_2010}, HAT-P-26\,b \citep{Hartman_etal_2011}, HAT-P-11\,b \citep{Yee_etal_2018}, and GJ\,3470\,b \citep{Kosiarek_etal_2019}.
In fact, among all Neptune-size planets with $P_{\rm orb} < 5$~day (Fig.~\ref{HNFig}), only HD\,219828\,b \citep{Melo_etal_2007} and HD\,47186\,b \citep{Bouchy_etal_2009} exhibit a small measured eccentricity ($e \lesssim 0.05$), which is still compatible with a nonzero value.


Figure~\ref{HNFig} shows measurements of eccentricity for planets with $P_{\rm orb} < 100$~day, divided into three groups: Earth-size ($R_\oplus < R < 3 \, R_\oplus$); Neptune-size ($3 \, R_\oplus < R < 9 \, R_\oplus$); and Jupiter-size ($9 \, R_\oplus < R < 25 \, R_\oplus$). 
The distribution of Neptune-size planets with $P_{\rm orb} < 5$~day contrasts with that given for close-in Jupiter- or Earth-size planets.
First, there are no warm Neptune planets with $P_{\rm orb} < 3$~day, 
which corresponds to the well-known ``Neptunian desert'' \citep{Lecavelier_2007, Davis_Wheatley_2009, Szabo_Kiss_2011, 
Mazeh_etal_2016}.
We note that the lack of eccentricity values for Earth-size planets is due to the difficulty in measuring them at a high precision level. 
Second, as noted above, all warm Neptune planets with $P_{\rm orb} < 5$~day have eccentricities that are consistent with nonzero values despite having damping timescales shorter than 5~Gyr. 
Finally, there is no apparent correlation between orbital period and eccentricity; on average, planets with shorter orbital periods are expected to have smaller eccentricities because they have shorter damping timescales (see Eq.\,(\ref{180702c})).

We observe that a fraction of Jupiter- and Earth-size planets with $P_{\rm orb} < 5$~day are also observed in eccentric orbits. 
Nevertheless, the eccentricity values increase with the orbital period, as expected, due to a more efficient tidal damping \bfx{at smaller distances} (Eq.\,(\ref{180702b})).
In the case of Jupiter-size planets, the eccentricity leftovers are likely related to their formation process through Lidov-Kozai cycles \citep[e.g.,][]{Fabrycky_Tremaine_2007} or planet-planet scattering \citep[e.g.,][]{Beauge_Nesvorny_2012}. 
Indeed, for $P_{\rm orb} > 5$~day, we observe that the eccentricity of Jupiter-size planets can reach extremely high values, while for the lower mass population it remains smaller than about 0.2.

One possibility to explain the observed eccentricities for warm Neptunes is that we are overestimating the tidal dissipation for these planets. 
If we adopt $Q/k_2 > 10^7$, the present nonzero eccentricities can be simply explained as residual values excited during the formation process.
However, such high values for $Q$ are not observed for similar mass planets in our Solar System \citep[e.g.,][]{Tittemore_Wisdom_1990, Banfield_Murray_1992}. 
In addition, this explanation also fails to explain the ``Neptunian desert'' and the absence of correlation between orbital period and eccentricity for $P_{\rm orb} < 5$~day (Sect.~\ref{hiecc}). 
Another possibility to explain the anomalous eccentricity distribution is that additional mechanisms pump the eccentricity or delay the circularization of the orbit.
In this paper, we discuss the most likely mechanisms that act in opposition to gravitational tides, namely, thermal atmospheric tides (Sect.~\ref{atides}), evaporation of the atmosphere (Sect.~\ref{evaporation}), and excitation from a distant companion (Sect.~\ref{companion}). We discuss their limits of applicability in Sect.~\ref{disccon}.

\section{High initial eccentricity}

\label{hiecc}

The damping timescale strongly depends on the orbital period, $\tau_\mathrm{circ} \propto P_\mathrm{orb}^{13/3}$  (Eq.~(\ref{180702c})).
Then, if the observed eccentric planet was initially on a wider and more eccentric orbit,
the initial damping timescale is 
much longer than the $\tau_\mathrm{circ}$ obtained with the current orbital period. 
In this section, we estimate this effect.

\begin{figure}
  \centering
  \includegraphics[width=\columnwidth]{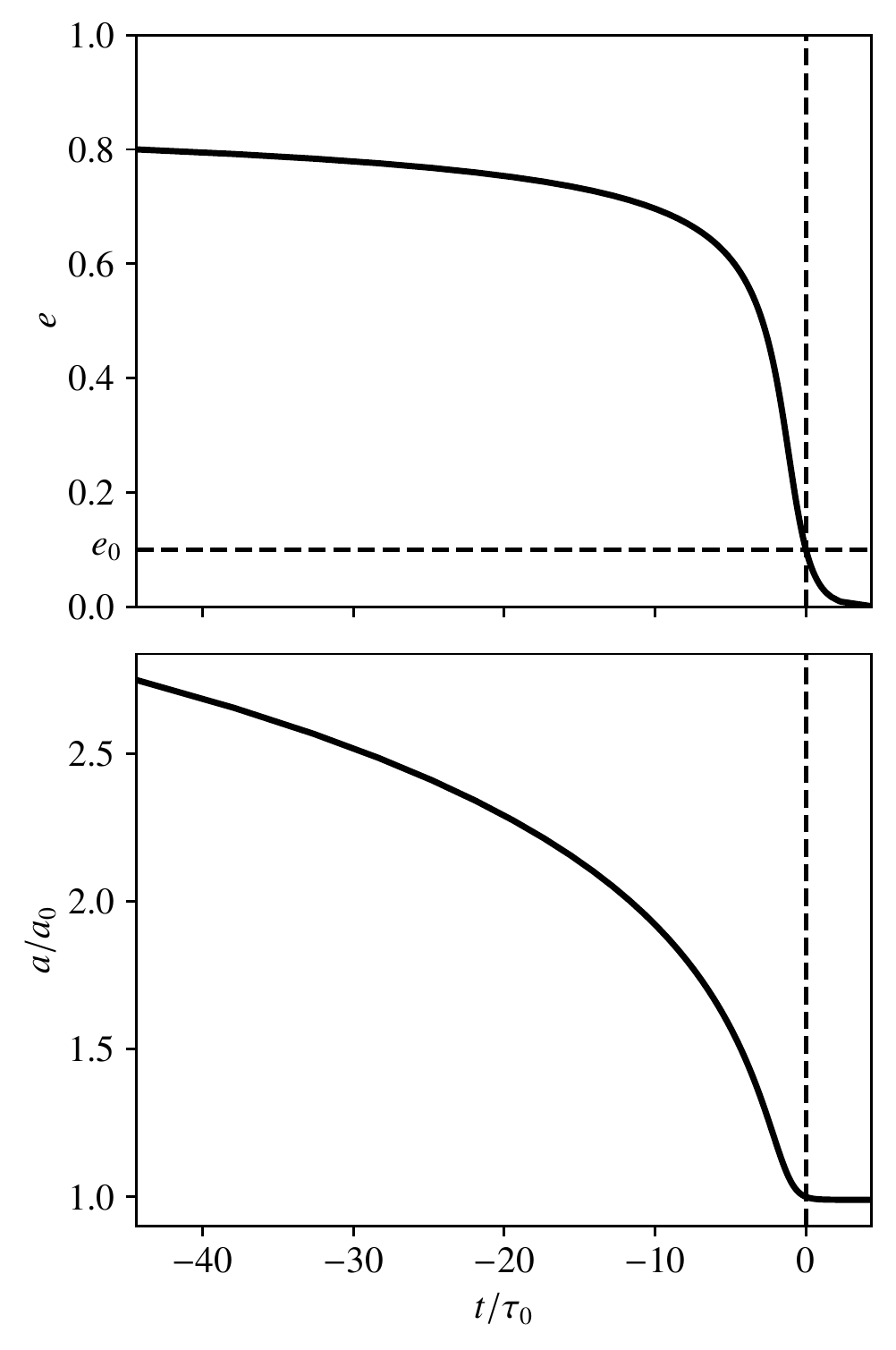}
  \caption{Evolution (past and future) of the eccentricity (\textit{top}) and semi-major axis (\textit{bottom})
    of a planet undergoing tidal dissipation with a power law $\dot{e}\propto - e a^{-13/2}$ (i.e. $\alpha=13/2$).
    The timescale $\tau_0$ is the currently observed instantaneous damping timescale (in this example at $a=a_0$, $e=e_0=0.1$).
    However, the shape of the curve is always the same and does not depend on $e_0$.}
  \label{fig:evolae}
\end{figure}

We assume that tides in the planet circularize its orbit following the law (see Eq.\,(\ref{180501a}))
\begin{equation}
  \dot{e} = -\frac{e}{\tau_\mathrm{circ}} \ , 
\end{equation}
with
\begin{equation}
    \tau_\mathrm{circ}(a) = \tau_0 \left(\frac{a}{a_0}\right)^\alpha \ .
\end{equation}
\bfx{Adopting the constant-$Q$ model\footnote{\bfx{For close-in planets, the orbital mean motion does not change much during the circularization process (see Fig.~\ref{fig:evolae}), so the constant-$Q$ model is a good approximation. Alternative tidal models provide different $\alpha$, which does not modify the final conclusions of this section.}} (Eq.\,(\ref{180702b}))}, we have $\alpha=13/2$.
We also assume that the tidal effect in the planet does not affect its orbital angular momentum significantly.
We thus have
\begin{equation}
  a \left(1-e^2\right) = a_0 \left(1-e_0^2\right) 
\end{equation}
and
\begin{equation}
  \label{eq:dote}
  \dot{e} = -\frac{e}{\tau_0} \left(\frac{1-e^2}{1-e_0^2}\right)^\alpha \ ,
\end{equation}
where $a_0$, $e_0$, and $\tau_0$ are the current (at the time of observation) values of the parameters.
In the case of small eccentricity, this equation can be approximated by
\begin{equation}
  \dot{e} = -\frac{e}{\tau_0},
\end{equation}
which is equivalent to assuming a constant circularization timescale $\tau_\mathrm{circ}$.
The evolution of the eccentricity in this regime is the classical exponential decay,
\begin{equation}
  \label{eq:evosmalle}
  e = e_0 \mathbf{e}^{-t/{\tau_0}}.
\end{equation}
On the contrary, if the eccentricity is close to one, expression (\ref{eq:dote}) can be approximated by
\begin{equation}
  \dot{x} = \frac{2}{\tau_0} \left(\frac{x}{x_0}\right)^\alpha,
\end{equation}
with $x=1-e^2$, and we find
\begin{equation}
  \label{eq:evobige}
  \frac{t}{\tau_0} = \frac{x_0}{2(\alpha-1)} \left(1-\left(\frac{x_0}{x}\right)^{\alpha-1}\right).
\end{equation}
Therefore, asymptotically (for $e\rightarrow 1$), we have $t \rightarrow -\infty$.
This means that we can make the tidal circularization process arbitrarily long by taking an initial eccentricity very close to one, together with a very large initial semi-major axis.

In the general case, the evolution of the eccentricity follows
\begin{equation}
  \frac{t}{\tau_0} = - \left(1-e_0^2\right)^\alpha \int_{e_0}^{e(t)} \frac{\mathrm{d}e}{e\left(1-e^2\right)^\alpha},
\end{equation}
which can be rewritten as
\begin{equation}
  \label{eq:evoe}
  \frac{t}{\tau_0} = \frac{1-e_0^2}{2(\alpha-1)} \left(
    f_\alpha(e_0) - \left(\frac{1-e_0^2}{1-e^2}\right)^{\alpha-1}f_\alpha(e)
    \right) \ ,
\end{equation}
where
\begin{equation}
  f_\alpha(e) = {}_2F_1\left(1,1-\alpha,2-\alpha,1-e^2\right)
\end{equation}
is the hypergeometric function.
The above expression allows us to estimate the time needed to reach the currently observed
orbit ($a_0$, $e_0$) as a function of the initial eccentricity $e$.
Figure~\ref{fig:evolae} shows an example of the evolution of the semi-major axis and eccentricity as a function of time.
We assume the observed eccentricity to be $e_0=0.1$, typical of the values measured for warm Neptunes (Fig.~\ref{HNFig}).
As expected, for small eccentricities ($e\lesssim 0.4$), we observe an exponential decay of the eccentricity (Eq.\,(\ref{eq:evosmalle})),
while at high eccentricities ($e\gtrsim 0.6$) the evolution is much slower (Eq.\,(\ref{eq:evobige})).
This counter-intuitive behaviour is due to the fact that when the eccentricity was large,
the semi-major axis was also large and, thus, tidal effects were much less efficient.

In Fig.~\ref{fig:evolae}, we see that if the initial eccentricity was around 0.8, and the initial semi-major axis around 2.8 times the observed one,
the planet might have spent around 40 $\tau_0$ to reach the observed configuration (where $\tau_0$ is the current instantaneous damping timescale). 
For a warm Neptune like GJ\,436\,b, this corresponds to about 8\,Gyr. 
Therefore, any mechanism (scattering, Kozai, tidal pumping, etc.) that could have excited the eccentricity to a high level during the formation of the planet (or at a lower level more recently) might be responsible for the currently remaining eccentricity. 

A possible limitation of this scenario is that we have to observe the system at a specific time in its evolution. 
Indeed, the eccentricity spends a long time above 0.6, and a long time close to 0, but a short time in-between (see Fig.~\ref{fig:evolae}).
The probability to observe the system at this specific time (i.e. during the phase of exponential decay) is about $\tau_0/T$, where $T$ is the star age. 
Therefore, if the observed population of warm Neptunes with eccentricity of $\sim$0.1 formed this way (Fig.~\ref{HNFig}), one would expect to also observe sizeable populations of Neptune-mass planets in the high eccentricity phase (on wider orbits) and in the circular phase (on closer-in orbits).

\section{Thermal atmospheric tides}
\llabel{atides}

The differential absorption of the stellar heat by the atmosphere of a planet gives rise to local variations of temperature and, consequently, to pressure gradients. 
The mass of the atmosphere is then permanently redistributed, adjusting for an equilibrium position.
More precisely, the particles of the atmosphere move from the high temperature zone (at the substellar point) to the low temperature areas.
Observations on Earth show that the pressure redistribution is essentially a superposition of two pressure waves \citep{Chapman_Lindzen_1970, Auclair-Desrotour_etal_2017a}: a daily (or diurnal) tide of small amplitude (the pressure is minimal at the subsolar point and maximal at the antipode) and a strong half-daily (semi-diurnal) tide (the pressure is minimal at the subsolar point and at the antipode).

As for bodily tides, there is a delay between the response of the atmosphere and the thermal excitation.
The phase shifted elongation of the shell induces a torque which modifies the rotational dynamics of the body as well as its orbit.
For atmospheric tides, the dominating semi-diurnal pressure wave usually leads the perturbation.
As a consequence, the thermal tide moves the planet rotation away from the synchronous equilibrium, and determines possible new states of equilibrium balanced by bodily tides \citep{Gold_Soter_1969, Correia_Laskar_2001}.

The combined effect of bodily and thermal tides tend to align the equator of the planet with the orbital plane \citep{Correia_etal_2003}.
Therefore, for simplicity, we let the obliquity of the planet to be zero.
In this case, the secular evolution of the rotation rate, $\omg$, and semi-major axis, $a$, due to a given kind of tidal effect (bodily or thermal) can be written for small eccentricity, $e$,
as \citep{Correia_Laskar_2003JGR, Correia_Laskar_2010B}:
\be
   \frac{d \omg}{d t} = - \frac{3}{2 C} K_\tau \, b_\tau (2 \omg - 2 n) 
   \ , \llabel{021010c1} 
\ee
\be
   \frac{d a}{d t} = \frac{3 a}{L} K_\tau \, b_\tau (2 \omg - 2 n) 
   \ , \llabel{100208a} 
\ee
where terms in $e^2$ have been neglected. 
$C$ is the principal moment of inertia, $L = \beta n a^2$ is the orbital angular momentum, $n$ is the mean motion, and $\beta = m M / (m+M) \approx m$ is the reduced mass of the system.
$K_\tau$ is a constant factor related to the strength of the tide, and $b_\tau(x)$ is an odd function related to the dissipation within the planet.
For bodily (or gravitational) tides we denote $\tau \equiv g$, while for thermal (atmospheric) tides we denote $\tau \equiv a$.
For the first kind we have
\be
K_g = \frac{G M^2 R^5}{a^6} \quad \mathrm{and} \quad b_g (x) = k_2 (x) \sin \delta_g(x) 
\llabel{021010e} \ , 
\ee
while for the second kind
\be
K_a = \frac{4 \pi M R^6}{5 m a^3} \quad \mathrm{and} \quad b_a (x) = - p_2 (x) \sin \delta_a(x) 
\llabel{VV9}  \ ,
\ee
where $G$ is the gravitational constant.
$k_2(x) > 0 $ is the second Love number, $p_2 (x) > 0$ is the amplitude of the pressure variations at the ground, and $0 \le \delta_\tau(x) < 90^\circ $ is the phase shift. 
The minus sign in the expression of $b_a(x)$ already accounts for the fact that thermal tides lead the perturbation \citep{Correia_etal_2003, Auclair-Desrotour_etal_2017a}.
The phase shift can be related with the time delay $\Delta t_\tau$ between the perturbation and the maximal amplitude of the tide through $\delta_\tau(x) = x \Delta t_\tau(x)$.
We thus have $b_\tau(0) = 0$, and for $x>0$, $b_g (x) > 0$ and $b_a(x) < 0$.
The phase shift can also be related to the quality factor through $Q_\tau^{-1} (x) = \sin \delta_\tau (x) $ \citep[][Eq.\,(141)]{Efroimsky_2012}.

For the eccentricity evolution, bodily and thermal tides give different evolutions\footnote{The bodily tides potential is proportional to $r^{-6}$, while the thermal tides potential is proportional to $r^{-5}$, where $r$ is the relative distance between the planet and the star \citep[see][]{Correia_etal_2003}.}. 
For the first we have
\be 
\begin{split}                 
   \frac{d e}{d t} = - \frac{3 e}{16 L} K_g &
   \Big[ 6 \, b_g (n) + 4 \, b_g (2\omg-2n)  \\ &
   + \, b_g (2\omg-n)  - 49 \, b_g (2\omg-3n) \Big] 
\ , \llabel{100208b} 
\end{split}
\ee
while for the second
\be 
\begin{split}                 
   \frac{d e}{d t} = - \frac{3 e}{16 L} K_a &
   \Big[ 4 \, b_a (n) + 4 \, b_a (2\omg-2n)  \\ &
   + 2 \, b_a (2\omg-n)  - 42 \, b_a (2\omg-3n) \Big] 
\ , \llabel{121208a} 
\end{split}
\ee

If we consider bodily tides alone, from equation (\ref{021010c1}) we see that the equilibrium configuration $\dot \omg = 0$ is obtained when for $\omg = n$, that is, for synchronous rotation.
In addition, this equilibrium is stable because $b_g(x)$ is an odd function for which $b_g (x) > 0$ when $x>0$.
Replacing $\omg = n$ in equations (\ref{100208a}) and (\ref{100208b}), we get $\dot a =0$, and 
\be
\frac{d e}{d t} = - \frac{21 e}{2 L} K_g b_g(n) < 0 \ ,
\llabel{180501a}
\ee
which leads to an exponential damping of the eccentricity, as expected.
For thermal tides alone, we can derive similar conclusions, but $\omg = n$ becomes an unstable equilibrium and (Eq.\,(\ref{121208a}))
\be
\frac{d e}{d t} = - \frac{9 e}{L} K_a b_a (n)  > 0 \ ,
\llabel{121208b}
\ee
since $b_a (x) < 0$ when $x>0$.
Hence, we conclude that thermal tides may be able to counterbalance the damping effect of bodily tides on the eccentricity.

When bodily and thermal tides are considered together, the complete evolution of the rotation rate  is given by (Eq.\,(\ref{021010c1}))
\be
   \frac{d \omg}{d t} = - \frac{3}{2 C} \left[ K_g \, b_g (2 \omg - 2 n) + K_a \, b_a (2 \omg - 2 n) \right]   
\ . \llabel{180501b} 
\ee
New equilibrium solutions for the rotation rate ($\dot \omg = 0$) are then obtained for \citep{Correia_Laskar_2001}
\be
f(2 \omg - 2 n) = - \frac{K_g}{K_a}  \ , \llabel{180501c} 
\ee
where $f(x) = b_a (x)/b_g (x)$.
This simple condition explains why the spin of the planet Venus is not synchronous.
Let us express the stable solutions of equation (\ref{180501c}) as  $\omg = n + \omg_0 /2 $. 
For the stable states 
we then have $f(\omg_0) = - K_g/K_a$.
Thus, for the semi-major axis secular evolution we get (Eq.\,(\ref{100208a}))
\be
   \frac{d a}{d t} = \frac{3 a}{L} \left[ K_g \, b_g (\omg_0) + K_a \, b_a (\omg_0) \right] = 0
   \ , \llabel{180501d} 
\ee
and for the eccentricity (Eqs.\,(\ref{100208b}) and\,(\ref{121208a}))
\be 
\begin{split}                 
   \frac{d e}{d t} = - \frac{3 e}{16 L} &
   \Big[ K_g \Big( 6 \, b_g (n) + b_g (n+\omg_0) + 49 \, b_g (n-\omg_0) \Big) \\ &
   + K_a \Big( 4 \, b_a (n) + 2 \, b_a (n+\omg_0) + 42 \, b_a (n-\omg_0) \Big) \Big]
\ . \llabel{180501e} 
\end{split}
\ee

The eccentricity evolution depends on the balance between the bodily and thermal tides harmonic functions $b_\tau(x)$, that is, it depends on the tidal models that we adopt.
As the rheology of planets is poorly known, these functions are usually unknown.
However, for close-in planets it is expected that bodily tides dominate thermal tides \citep{Correia_etal_2008, Cunha_etal_2015, Leconte_etal_2015} and, thus, $|\omg_0| \ll n$. 
We can then simplify the previous expression as
\be 
\begin{split}                 
   \frac{d e}{d t} & \approx - \frac{3 e}{2 L} \Big[ 7 \, K_g b_g (n) + 6 \, K_a b_a (n)  \Big]  \\ &  
   = - \frac{21 e}{2 L} K_g  \Big[ b_g (n) - \frac67 \frac{b_a (n)}{f(\omg_0)}  \Big]
\ , \llabel{180501g} 
\end{split}
\ee
that is, the eccentricity increases whenever
\be
\frac{b_a (\omg_0)}{b_a (n)} <  \frac67 \frac{b_g (\omg_0)}{b_g (n)} \ . 
\llabel{181215a} 
\ee

We immediately see that the eccentricity always decreases if we adopt for both tides a constant ($b_\tau = cte$) or a linear model ($b_\tau \propto x$).
On the other hand, if we adopt a linear model for thermal tides and a constant model for bodily tides, the eccentricity always increases, since we get $\omg_0/n < 6/7$, which is consistent with $|\omg_0| \ll n$.
The constant and linear models are widely used, because they are simple, but they are not very realistic.
A more correct description of the rheology involves the use of viscoelastic models \citep[e.g.,][]{Henning_etal_2009}.
In particular, thermal tides are believed to follow a Maxwell rheology \citep{Leconte_etal_2015,Auclair-Desrotour_etal_2017a}, for which $b_a(x) \propto - x / (1 + \tau_a^2 x^2)$, where $\tau$ is a viscous relaxation time.
Assuming also a Maxwell model for bodily tides, with $b_g(x) \propto x / (1 + \tau_g^2 x^2)$ \citep{Remus_etal_2012a, Correia_etal_2014}, the eccentricity increases for (Eq.\,(\ref{181215a}))
\be
\frac{\omg_0}{n} \frac{1+\tau_a^2 n^2}{1+\tau_a^2 \omg_0^2} 
< \frac{\omg_0}{n} \frac{1+\tau_g^2 n^2}{1+\tau_g^2 \omg_0^2} \ ,
\llabel{191009a} 
\ee
that is, for
\be
\tau_g > \tau_a \quad \mathrm{and} \quad |\omg_0| < n \ .
\llabel{181215b} 
\ee
The value of $\tau_g$ can be estimated from $Q \sim \tau_g n + 1/(\tau_g n) $, which has two solutions. 
With $Q\sim 10^4$, for orbital periods $P_{\rm orb} \sim 5$~day, we get $\tau_g \sim 10^4$~day or $\tau_g \sim 10^{-4}$~day.
At present, there are no estimations of $\tau_a$ for Neptune-like planets.
However, for an Earth-like planet, we get $\tau_a \sim 10$~day \citep{Leconte_etal_2015}.
Assuming an identical value for warm Neptunes, we see that condition (\ref{181215b}) is widely assured for $\tau_g \sim 10^4$~day.
We then conclude that although there is a limited set of values $(\tau_g, \tau_a)$ that allow an increase in the eccentricity due to the combined action of bodily and thermal tides, the rheology of warm Neptunes may be such that condition (\ref{181215b}) can be met.

\section{Evaporation of the atmosphere}
\llabel{evaporation}

Interestingly, it has been observed that the atmospheres of some of the warm Neptunes with nonzero eccentricity are also undergoing strong evaporation. This is the case for the planets GJ\,436\,b \citep{Ehrenreich_etal_2015, Lavie_etal_2017, dosSantos_etal_2019} and GJ\,3470\,b \citep{Bourrier_etal_2018a}. Therefore, a fraction of the mass of these planets is being lost through this process. Isotropic planetary evaporation generates almost no orbital evolution. However, the hottest region in a planet atmosphere could shift with respect to the substellar point, usually eastward, depending on the strength of the imposed stellar heating and other factors such as the radiative timescale and drag time constant \citep[e.g.,][]{Showman_Guillot_2002, Showman_Polvani_2011}.
This displacement could be associated with anisotropic mass-loss, and a modification in the orbit of the planet \citep{Boue_etal_2012, Teyssandier_etal_2015}.

We let $\theta$ be the angle between the star and the direction of maximal mass ejection, $\varphi$ be the solid angle aperture of the stream, and $v_e$ be the ejection speed of atmospheric particles with respect to the planet's barycenter \citep[see Fig.\,1 in][]{Boue_etal_2012}.
The secular evolution of the semi-major axis and eccentricity due to the mass-loss can be written as \citep{Boue_etal_2012}:
\be
   \frac{d a}{d t} = - \frac{\dot m}{m} \left(  2 \gamma + \frac{m}{M} \right) a
   \ , \llabel{180710a} 
\ee
\be 
   \frac{d e}{d t} = - \frac{\dot m}{m} \frac{\gamma}{2} e
   \ , \llabel{180710b} 
\ee
where terms in $e^2$ have been neglected, and
\be
\gamma = \sin \theta \cos^2 \frac{\varphi}{2} \left(\frac{v_e}{n a}\right)  \llabel{180710c} 
\ee
is a dimensionless parameter related to the efficiency of the evaporation.
We have $\gamma=0$ for substellar ($\theta=0$, $\pi$) or isotropic evaporation ($\varphi=\pi$).
We also note that the orbital speed $v_\mathrm{orb} \approx n a$. 
Since we assume that the planet is losing mass ($\dot m < 0$), this leads to an increase in both semi-major axis and eccentricity.

The secular evolution of the eccentricity due to the combined effect of bodily tides (Eq.\,(\ref{180501a})) and evaporation (Eq.\,(\ref{180710b})) is then
\be 
   \frac{d e}{d t} = - \left[ \frac{21}{2 L} K_g b_g(n) + \frac{\dot m}{m} \frac{\gamma}{2} \right] e
   \ , \llabel{180710d} 
\ee
which is positive for
\be 
- \frac{\dot m}{m} \, \gamma >  \frac{2}{\tau_\mathrm{circ}}
   \ . \llabel{180710e} 
\ee

For GJ\,436\,b and GJ\,3470\,b we need $- \gamma \, \dot m / m > 10$~Gyr$^{-1}$ and $3.3$~Gyr$^{-1}$, respectively (Eq.\,(\ref{180702c})). 
The two planets have similar orbital velocities, $v_\mathrm{orb} \approx 120$~km\,s$^{-1}$. 
Theoretical simulations of escaping outflows yield typical escape velocities $v_e$ between $1-10$~km\,s$^{-1}$ \citep[e.g.,][]{Salz_etal_2016a}. Simulations of GJ\,436\,b exosphere based on Lyman-$\alpha$ transit observations suggest that $v_e$ could reach up to $50-60$~km\,s$^{-1}$. 
Adopting the upper limit, $\gamma$ still remains lower than about 1. 
The present mass loss rates from GJ\,436\,b and GJ\,3470\,b atmospheres are $2.2\times10^{10}$~g\,s$^{-1}$ \citep{Bourrier_etal_2016} and $8.5\times10^{10}$~g\,s$^{-1}$ \citep{Bourrier_etal_2018a}, respectively. 
These values are also upper limits, assuming the planets' atmospheres are in the energy-limited regime and all input stellar energy is used for the escape. 
Even with maximum values, $- \gamma \, \dot m / m$ would thus reach 0.0025~Gyr$^{-1}$ for GJ\,436\,b and 0.016~Gyr$^{-1}$ for GJ\,3470\,b, far too low for anisotropic escape to balance the damping effect from tides. 
Indeed, in the optimistic case for which $\gamma \sim 1$, the planet would have to lose roughly half of its mass every 100~Myr to balance the damping effect from tides;
after a few Myr the planet would have completely lost its atmosphere.
We then conclude that, unless there is an unknown mechanism that produces an ejection speed of $v_e > 10^3$~km\,s$^{-1}$, the orbital effect of atmospheric evaporation is not sufficient to explain the eccentricities observed for warm Neptunes.

\section{Excitation from a distant companion}
\llabel{companion}



Most warm Neptune eccentric planets were discovered using radial velocity combined with transit observations.
At present, these are the two most successful techniques for detecting planets, but they share a caveat: they are unable to spot planets whose orbital plane lie near the observer's plane of the sky.
In the case of radial velocity, the amplitude of the signal is proportional to the sinus of the inclination of the orbital plane to the plane of the sky, while for transits we can only detect planets passing in front of the star.
Therefore, we cannot rule out that a companion is present in the system, provided that it lies in a inclined or very distant orbit with respect to the warm Neptune.
Indeed, some warm Neptune planets have been observed in systems with more massive distant companions (HAT-P-11, HD\,219828, HD\,47186 and HD\,125612).
We note, however, that even in these positive cases, we also lack knowledge for their true masses and inclinations, since they were all spotted by radial-velocity alone.
In other cases, such as GJ\,436, GJ\,3470 and HAT-P-26, no companion has been identified so far. 

In planetary dynamics, still undetected companions are often suggested to solve unexplained observations.
For instance, a ninth planet was proposed to explain the observed clustering in the perihelia of the distant Kuiper Belt objects \citep{Trujillo_Sheppard_2014, Batygin_Brown_2016}.
A great success was the discovery of the planet Neptune by \citet{LeVerrier_1846}, that probably triggered all the other undetected companion scenarios.
A great failure was the prediction of planet {\it Vulcan} to explain the perihelion advance of Mercury, also by \citet{LeVerrier_1859}, which was later explained in the frame of general relativity \citep{Einstein_1915}.
Therefore, the undetected companion scenario should always be considered with a degree of caution.

When the orbit of a planet is excited by an outer companion in a inclined orbit, its eccentricity undergoes some perturbations.
Assuming for simplicity that the perturber is on a circular orbit with radius $a_p$, mutual inclination $I$, and its mass $m_p \gg m$, the dominating (quadrupolar) contribution for the eccentricity and pericenter are \citep[e.g.,][]{Correia_etal_2013}:
\be
\dot e \approx  \tfrac{5}{2} \nu \, e \, (1-e^2)^{1/2}  \sin^2 I \sin 2 \vpi \ , \llabel{120621a}
\ee
\be 
\begin{split}  
\dot \vpi \approx & \frac{\nu_g}{ (1-e^2)} + \frac{\nu_r \, (\omg/n)^2}{(1-e^2)^2} \\ &
+ \nu  \frac{2 (1-e^2) + \frac{5}{2} (e^2 - \sin^2 I) (1 - \cos 2 \vpi)}{(1-e^2)^{1/2} } 
\ , \llabel{180713a} 
\end{split}
\ee
where
\be
\nu =  n \frac{3}{4} \frac{m_p}{M} \left(\frac{a}{a_p}\right)^3 \ , \llabel{110817b}
\ee
is due to the perturber, while 
\be
\nu_g = 3 n \left(\frac{n a}{c}\right)^2 
\, \quad \mathrm{and} \quad
\nu_r = n \frac{k_2}{2} \frac{M}{m} \left(\frac{R}{a}\right)^5
 \ , \llabel{110817a}
\ee
are due to general relativity and planet oblateness, respectively.
The mutual inclination can also change over time, but it can be related with the eccentricity using the conservation of the total orbital angular momentum: 
\be
h = \sqrt{1-e^2} \cos I \approx cte 
\ , \llabel{180713z}
\ee
also known by the ``Kozai constant.''

\subsection{Lidov-Kozai cycles}

When a massive companion in a inclined orbit is present, there is an alternative mechanism to the migration in the accretion disk for the formation of close-in planets: 
high eccentricity migration through Lidov-Kozai cycles combined with tidal friction \citep[e.g.,][]{Wu_Murray_2003, Correia_etal_2011, Anderson_etal_2016}. 
This mechanism has been proposed by \citet{Beust_etal_2012} to explain the present eccentricity of GJ\,436\,b and it is strengthened by the measurement of the planet misaligned orbit \citep{Bourrier_etal_2018b}, which is also a natural consequence of the Kozai cycles.

For planets far from the star, the gravitational perturbations from the companion dominate over the general relativity and the oblateness terms ($\nu \gg \nu_g$, $\nu_r$) in the precession of the periapsis (Eq.\,(\ref{180713a})). 
As a consequence, it is possible to find stable equilibria points for the eccentricity ($\dot e = 0$, $\dot \vpi = 0$), known as Kozai equilibria \citep{Lidov_1962, Kozai_1962}.
These points are centered at $\vpi = \pm \pi/2$ and $\cos I = \sqrt{3/5} \sqrt{1-e^2}$.
They correspond to a resonance between the precession of the longitude of the node and the longitude of the pericenter of the inner orbit.
Therefore, there is a libration zone associated to this resonance (for $\cos I < \sqrt{3/5}$), a separatrix, and a non-resonant circulation zone.

High-eccentricity migration through Lidov-Kozai cycles is very efficient, provided that the planet is close to the separatrix of the Lidov-Kozai equilibria.
Indeed, near the separatrix, the mutual inclination undergoes large variations, that, in turn, induce large eccentricity oscillations (Eq.\,(\ref{180713z})).
Even at large distances, the outer companion can significantly perturb the inner orbit as long as the initial mutual inclination is $\cos I < \sqrt{3/5}$.
If the inner orbit is initially circular, the maximum eccentricity achieved in a Lidov-Kozai cycle is $ e_{max} = \sqrt{1 - (5/3) \cos^2 I} $ \citep[e.g.,][]{Lidov_Ziglin_1976}.
When the eccentricity reaches very high values tidal dissipation is enhanced and the semi-major axis of the inner orbit decreases.

Lidov-Kozai cycles persist as long as the perturbation from the outer companion is the dominant cause of periapse precession in the planetary orbit.
However, small additional sources of periapse precession, such as general relativity ($\nu_g$) or oblateness ($\nu_r$), can compensate the gravitational precession ($\nu$) and suppress the eccentricity oscillations (Eq.\,(\ref{120621a})). 
The inner planet is then left in a close-in orbit with large eccentricity that is then progressively damped by tides (see Sect.~\ref{hiecc}).

The presence of the Lidov-Kozai cycle from an undetected companion then act as possible mechanism of formation of the warm Neptune that simultaneously accounts for the possible observed eccentricity, as a remnant of this formation mechanism.
Although the circularization timescale for GJ\,436\,b is only about 0.2 Gyr (Eq.\,(\ref{180702c})), there is a range of parameter values for its companion for which the Lidov-Kozai cycles could delay the deliver of the planet close to the star; that would explain why we still observe $e \approx 0.16$ at present \citep{Bourrier_etal_2018b}.

\subsection{Spin-driven eccentricity pumping}

This effect has been first described for coplanar orbits \citep{Correia_etal_2012, Greenberg_etal_2013}, but it is also much more efficient for mutually inclined orbits \citep{Correia_etal_2013}.

When the general relativity or the oblateness terms dominate over the gravitational perturbations ($\nu_g$ or $\nu_r \gg \nu$), the Lidov-Kozai cycles no longer work.
This is the case for all currently observed warm Neptunes.
Thus, at first order, the precession of the periastron is constant, $ \dot \vpi \simeq g $ (Eq.\,(\ref{180713a})), and the eccentricity shows small oscillations around a mean value, $e = e_0 + \delta e$, with
\be
\delta e = \Delta e \cos  (g t + \vpi_0) \ , \llabel{110812e}
\ee
where $e_0$, $\Delta e$ and $\vpi_0$ are constant.
The evolution of the rotation rate (Eq.\,(\ref{021010c1})) also depends on the eccentricity,\footnote{In expression (\ref{021010c1}) we neglected terms in $e^2$, but expanding to a higher order we get \citep{Correia_etal_2008}: $C \dot \omg = - \frac{3}{2} K_g [ (1-5 e^2) b_g (2 \omg - 2 n) + \frac{1}{4} e^2 b_g (2 \omg - n) +\frac{49}{4} e^2 b_g (2\omg - 3 n) ] + {\cal O} (e^4)$.}
so the rotation rate experiences a variation $\omg = \omg_0 + \delta \omg$, with
\be
\delta \omg = \Delta \omg \cos ( g t + \vpi_0 - \phi) \ , \llabel{110920a}
\ee
where $\omg_0$, $\Delta \omg$ and $\phi$ are constant.
The rotation rate thus presents an oscillation identical to the eccentricity (Eq.\,(\ref{110812e})), but delayed by an angle $\phi$, such that \citep{Correia_etal_2013}
\be
\sin \phi \sim 2 g (k^2 + 4 g^2)^{-1/2} 
\ , \quad \mathrm{with} \quad 
k = \frac{K_g}{C n} \frac{k_2}{Q}
\ . \llabel{180712a}
\ee
The damping timescale for the spin can be seen as $\tau_\mathrm{spin} \approx k^{-1} $, which is usually much shorter than the circularization timescale, $\tau_\mathrm{circ} \approx L Q / k_2 K_g$ (Eq.\,(\ref{180702b})), since $L \gg C n$.
At second order, the precession of the periastron depends on the eccentricity (Eq.\,(\ref{180713a})), but also on the rotation rate (term in $\nu_r$).
Replacing expressions (\ref{110812e}) and (\ref{110920a}) in the expression of the periastron (Eq.\,(\ref{180713a})), integrating and replacing the result again in expression (\ref{110812e}), we obtain a secular contribution for the eccentricity \citep[see appendix A in][]{Correia_etal_2013}
\be
\delta \dot e \approx \frac{\nu_r}{8} \sin 2 \phi 
\ . \llabel{121217e}
\ee
Thus, the secular term vanishes when $\phi=0$ or $\pi/2$, that is, for strong dissipation ($k \gg g $), where $\delta \dot e \sim \nu_rg / k$, or for weak dissipation ($ k \ll g $), where $\delta \dot e \sim \nu_rk / g $, respectively.
The effect on the eccentricity is then maximized when $\phi = \pi/4$, which occurs for $  k \sim g $, for which $\delta \dot e \sim \nu_r$.
We conclude that, in order to observe the spin-driven eccentricity pumping effect, the damping timescale of the spin ($k^{-1}$) should be of the order of the period of eccentricity oscillations ($g^{-1}$).

The secular evolution that results from expression (\ref{121217e}) must be added to the usual orbital damping (Eq.\,(\ref{180501a})).
As long as the spin-driven term counterbalances the damping effect, the eccentricity can increase to high values.
However, when we consider the full non-linearized problem, the secular increase (Eq.\,(\ref{121217e})) cannot last indefinitely because when the eccentricity reaches high values, the damping effect is also enhanced.
As a consequence, the spin-driven pumping of the eccentricity is never permanent, although it can last through the age of the system.

%

\section{Discussion}
\label{disccon}

The distribution of eccentricity measured for warm Neptunes with $P_\mathrm{orb} < 5$~day contrasts with that given for smaller and larger cl ose-in planets. 
In particular, most warm Neptunes present nonzero eccentricity, a surprising feature considering that bodily tides should have circularized their orbits in a timescale shorter than the age of the systems. 
In this paper, we investigate several mechanisms that could counterbalance the bodily tides, namely, thermal atmospheric tides, evaporation of the atmosphere, and excitation from a distant companion.

The combined action of bodily and thermal tides can be simplified using their relaxation times, $(\tau_g, \tau_a)$, respectively.
A dynamical analysis shows that the eccentricity is allowed to increase under some conditions for these two parameters, in particular, for $\tau_a < \tau_g $ (Eq.\,(\ref{181215b})).
Observational constraints on the rheology and atmospheric composition of warm Neptunes are required to assess the fraction of these planets maintained on eccentric orbits because of thermal tides.

Spectacular evaporation has been observed for the warm Neptunes GJ\,436\,b \citep{Bourrier_etal_2018b} and GJ\,3470\,b \citep{Bourrier_etal_2018a}. 
This class of planets is expected to be particularly sensitive to atmospheric escape \citep{Owen_Wu_2017}. 
If anisotropic, their evaporation would always contribute to increase the orbital eccentricity. 
Our estimates, however, show that the velocity and mass loss rate of the gas escaping from warm Neptunes (as derived from observations and predicted theoretically) are not sufficient to counterbalance the damping effect from bodily tides.

Distant, more massive planets have been found to accompany some warm Neptunes. 
Those that appear alone may still have undetected companions. 
In that case, mutual gravitational interactions between the two planets can temporarily excite the eccentricity of the inner warm Neptune, thus delaying the circularization of the orbit up to several Gyrs.
This excitation can be induced via Lidov-Kozai cycles for companions with mutual inclinations $\cos I < \sqrt{3/5}$, or via spin-driven eccentricity pumping when the damping timescale of the spin ($k^{-1}$) is of the order of the period of eccentricity oscillations ($g^{-1}$). 
Because they excite the eccentricity for a limited amount of time, these mechanisms alone cannot explain the distribution of eccentricity observed for warm Neptunes, in particular the absence of close-in Neptunes on circular orbits.

Although all the proposed scenarios have some limitations, we cannot rule out that more than one of these mechanisms is simultaneously at work. 
It is possible that the hot expanding atmosphere of an evaporating Neptune creates an additional thermal tidal torque on the planet that would help to increase its eccentricity.
Alternatively, a distant companion can excite the eccentricity, and as the Neptune-size planet migrates closer to the star, it would evaporate increasingly and erode into a smaller planet before it can fully circularize, thus explaining why no Neptune-mass planets are found on circular orbits close to their star.
This combination of high-eccentricity migration and evaporation as the origin of eccentric warm Neptunes was proposed by \citet{Bourrier_etal_2018b}, based on the study of GJ\,436b, and is now gaining interest to explain the structure of the ``Neptune desert'' \citep[e.g.,][]{Owen_Lai_2018}.
In this scenario, the warm Neptunes on eccentric orbits 
are either still undergoing their migration and will erode as they enter the desert or they have already reached a stable orbit far enough from the star to be safe from evaporation.
The orbital properties and evaporation status of the warm Neptunes GJ\,436\,b \citep{Bourrier_etal_2018b} and GJ\,3470\,b \citep{Bourrier_etal_2018a} are consistent with this scenario.

In conclusion, thermal atmospheric tides or excitation from a distant companion combined with evaporation could explain the distribution of eccentricities observed for warm Neptunes. 
These scenarios require specific conditions for the warm Neptunes, their host star, and possible companions. 
At present we cannot conclude whether one of these mechanisms is dominant, whether they act together, or whether each influence a fraction of the observed warm Neptunes. 

Further theoretical modeling of these mechanisms should be attempted to better understand their impact, complemented by in-depth characterizations of the orbital and atmospheric properties of known warm Neptunes, along with a search for additional Neptune-size planets and their putative companions.
\bfx{Studying the distribution of eccentricities as a function of the tidal quality factor could also allow us to better assess the role of tidal circularization for each class of planets, though this requires a larger sample of high-precision mass and radius measurements, and a finer understanding of planetary internal structures.}

\begin{acknowledgements}
A.C. acknowledges support by 
CFisUC strategic project (UID/FIS/04564/2019),
ENGAGE SKA (POCI-01-0145-FEDER-022217), and
PHOBOS (POCI-01-0145-FEDER-029932),
funded by COMPETE 2020 and FCT, Portugal.
V.B. acknowledges support by the European Research Council (ERC) under the European Union's Horizon 2020 research and innovation programme (project Four Aces grant agreement No 724427).
J.-B.D. acknowledges support by the Swiss National Science Foundation (SNSF).
This work has, in part, been carried out within the framework of
the National Centre for Competence in Research PlanetS
supported by SNSF.
\end{acknowledgements}

\bibliographystyle{aa}
\bibliography{\bibpath correia}

\end{document}